\documentclass[conference]{IEEEtran}
\IEEEoverridecommandlockouts
\usepackage{cite}
\usepackage{amsmath,amssymb,amsfonts}
\usepackage{algorithmic}
\usepackage{graphicx}
\usepackage{textcomp}
\usepackage{xcolor}

\usepackage{hyperref}
\usepackage{pifont}
\newcommand{\cmark}{\textcolor{blue}{\ding{51}}}%
\newcommand{\xmark}{\textcolor{red}{\ding{55}}}%

\def\BibTeX{{\rm B\kern-.05em{\sc i\kern-.025em b}\kern-.08em
    T\kern-.1667em\lower.7ex\hbox{E}\kern-.125emX}}
\begin{document}

\title{Embedded Deployment of Semantic Segmentation in Medicine through Low-Resolution Inputs\\
}

\author{
    \IEEEauthorblockN{
    Erik Ostrowski\IEEEauthorrefmark{1}, Muhammad Shafique\IEEEauthorrefmark{3}}
    \IEEEauthorblockA{\IEEEauthorrefmark{1}Institute of Computer Engineering, Technische Universit{\"a}t Wien (TU Wien), Austria
    \\ erik.ostrowski@tuwien.ac.at}
    \IEEEauthorblockA{\IEEEauthorrefmark{3}eBrain Lab, Division of Engineering, New York University Abu Dhabi (NYUAD), United Arab Emirates (UAE)
    \\muhammad.shafique@nyu.edu}
}

\maketitle

\begin{abstract}
When deploying neural networks in real-life situations, the size and computational effort are often the limiting factors. 
This is especially true in environments where big, expensive hardware is not affordable, like in embedded medical devices, where budgets are often tight. 
State-of-the-art proposed multiple different lightweight solutions for such use cases, mostly by changing the base model architecture, not taking the input and output resolution into consideration. 
In this paper, we propose our architecture that takes advantage of the fact that in hardware-limited environments, we often refrain from using the highest available input resolutions to guarantee a higher throughput. 
Although using lower-resolution input leads to a significant reduction in computing and memory requirements, it may also incur reduced prediction quality. 
Our architecture addresses this problem by exploiting the fact that we can still utilize high-resolution ground-truths in training. 
The proposed model inputs lower-resolution images and high-resolution ground truths, which can improve the prediction quality by 5.5\% while adding less than 200 parameters to the model. %reducing the frames per second only from 25 to 20.
We conduct an extensive analysis to illustrate that our architecture enhances existing state-of-the-art frameworks for lightweight semantic segmentation of cancer in MRI images. 
We also tested the deployment speed of state-of-the-art lightweight networks and our architecture on Nvidia's Jetson Nano to emulate deployment in resource-constrained embedded scenarios.
%The framework is open-source and accessible online at \url{https://BlindedLinkForReview}. 
\end{abstract}


\section{Introduction}
Over time, AI has found many use cases in the medical field, like EEG analysis~\cite{ EEG2, EEG3}, improvement of MRI image resolution~\cite{SR2, SR3}, and detecting diseases in medical images. 
Above all, the detection and localization of cancer have gotten much attention from the AI research community.  
Hence, many methods were proposed for Computer Aided Diagnostics (CAD), starting exclusively with image classification in the early days \cite{CADC1, CADC2, CADC3} to the more fine-grained bounding boxes \cite{CADOD1, CADOD2, CADOD3}, and semantic segmentation predictions \cite{CADSS1, CADSS2, CADSS3}. 
Semantic Segmentation predictions consist of a pixel-wise classification for the input image, giving the most detailed degree of localization and, thus, the most helpful information for the doctors. 

The U-Net architecture has established itself as the most popular deep-learning model for semantic segmentation in the medical field. 
The U-Net architecture consists of an encoder and a decoder part. 
Given an input image, the encoder compresses its size multiple times while the decoder takes the compressed encoder input and decompresses it to the original size of the input. 
This architecture proved especially successful in medical applications since the network's encoding part efficiently captures the relevant information in the image, like most segmentation architectures. 
However, the subsequent decoding part up-samples the compressed input back to a higher resolution, which preserves the spatial information. 
%Furthermore, the authors of U-Net mention that their architecture performs well with few training images compared to other segmentation networks. 
In the medical field, we often deal with small datasets. 
\begin{figure}[t]
\centering
\includegraphics[width=0.9\columnwidth]{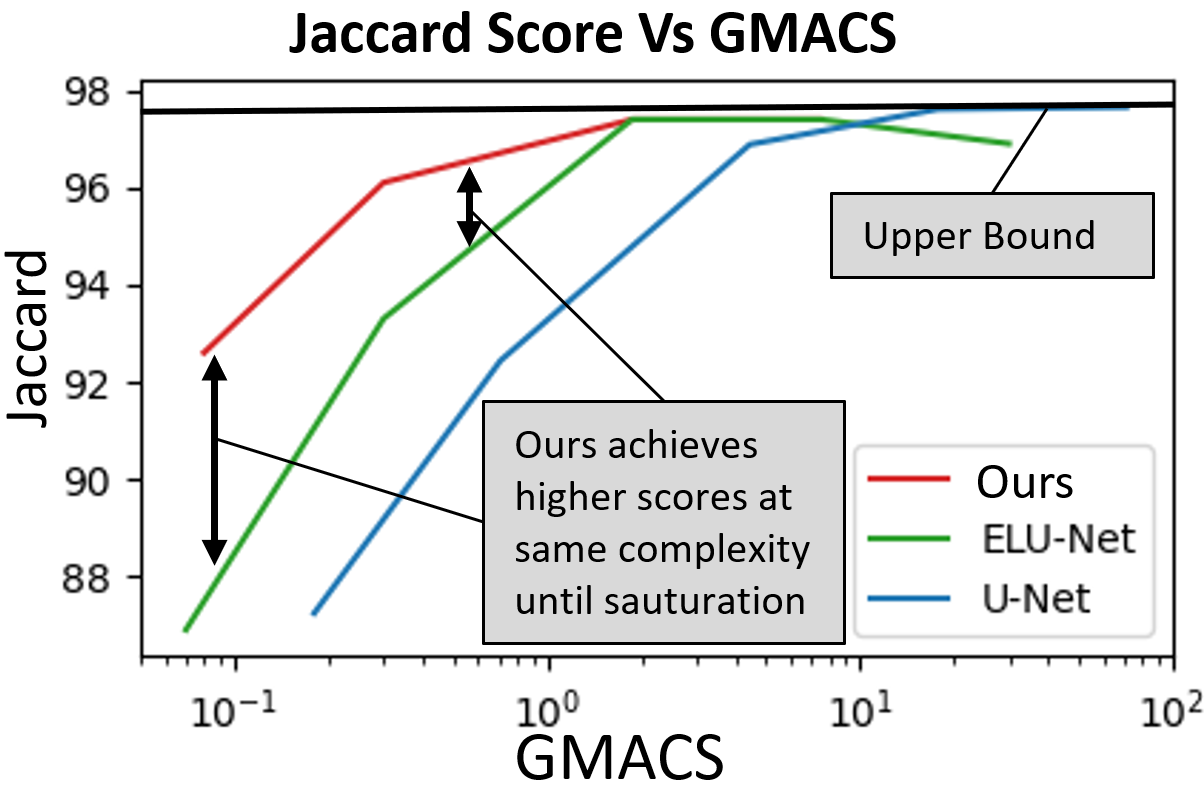} % width=50mm,scale=0.50 width=\linewidth
\caption{Trade-off between the number of Giga Multiply-Accumulate (GMAC) Operations and Jaccard prediction score of our architecture, U-Net~\cite{UNET} and state-of-the-art ELU-Net~\cite{ELUNET}
\label{graph1}}
\end{figure}
However, if we strive for real-world applicability, we can not focus only on prediction quality but also need to put emphasis on reducing the cost of application. 
On the one hand, it is convenient to make predictions on video streams, which require a throughput rate of at least highly one image per second and preferably around 20 images per second. 
On the other hand, being able to make predictions on small, efficient, and cheaper hardware is more than desired. 
%Adding the hardware limitations and inference speed requirements complicates things, but we can also fall back on the rich literature on lightweight networks for semantic segmentation. 
Research has already proposed several improvements and iterations on U-Net~\cite{UNET} like U-Net3+~\cite{UNET3+} or ELU-Net~\cite{ELUNET}, which partly also focus on throughput rate. 
In particular, ELU-Net achieved higher throughput at lower overall memory consumption while preserving or improving the original network's prediction quality. 
However, the proposed lightweight networks are often frigid. 
When they fail the set target, the end-user does not have much freedom to change the architecture to turn the dial more toward throughput speed at the cost of prediction quality. 
The go-to strategy in this scenario is to reduce the resolution of the input images. 
Reducing the input image resolution is very effective in meeting strict hardware limitations. 
For example, by reducing the resolution from $320 \times 320$ to $160 \times 160$, we reduce the memory consumption and number of operations by $75\%$.
However, with this comes also a considerable loss in prediction quality. 
To address this issue, we exploit two observations. 
First, most computations in a U-Net-like architecture stem from encoding the input image and not so much from the up-scaling in the decoder part of the network.  
Second, since we down-scaled the original input, we still have access to the high-resolution ground-truths. 
Hence, we propose our our architecture designed to take advantage of this scenario. 
Our architecture extends any U-Net-like architecture by adding more up-scaling layers at the end of the U-Net, therefore outputting a higher resolution than the input image. 
Those additional layers do not contribute much toward the overall computational complexity of the model but help the model to leverage the more informative high-resolution ground truth, thus improving the prediction quality of the network compared to only using low-resolution input and ground truth. 
To further boost the prediction quality, we propose a loss function using the output from all up-scaling steps, from the input resolution to the ground truth resolution. 
The loss values generated from the different up-scaling steps are helpful in guiding the model while not adding any significant computations. 
Fig.~\ref{novelty} compares the state-of-the-art approach to output the same resolution as the input, thus ensuring the dependability on expensive high resolution input for high quality results.
Whereas our architecture is still able generate meaningful predictions with very low resolution input. 

\begin{figure}[t]
\centering
\includegraphics[width=0.7\columnwidth]{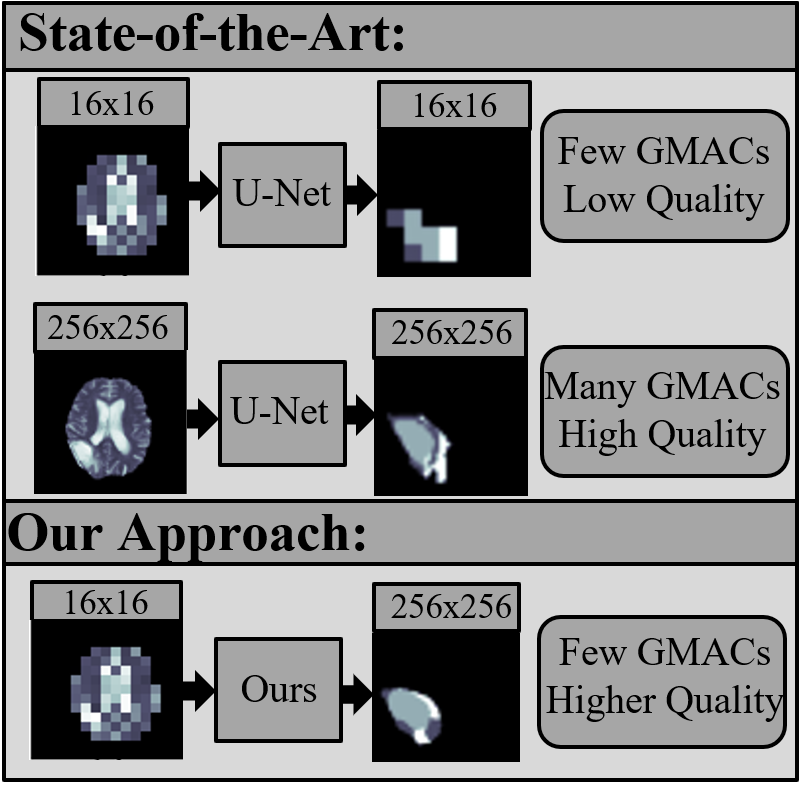} % width=50mm,scale=0.50 width=\linewidth
\caption{The novelty of our architecture lies in its ability to generate high resolution predictions with very compressed inputs.
\label{novelty}}
\end{figure}

Our experiments have shown that our architecture dramatically improves the prediction quality of the baseline when using the same input resolution while not adding any significant computations.
Fig.~\ref{graph1} shows the trade-off in prediction quality and number of Giga Multiply–Accumulate (GMAC) operations for a standard U-Net, ELU-Net, and our architecture.
As we increase the input resolution, the networks perform more GMACs but also achieve better prediction scores.
However, the standard U-Net reaches the upper bound only at very high input resolutions, while ELU-Net and our architecture reach the upper bound earlier.
If we take a look at the prediction scores before saturation, we notice that our architecture reaches higher prediction scores with less model complexity.
Moreover, we compared our architecture's performance to comparable U-Net architectures on Nvidia's Jetson Nano to show its viability in resource-constraint embedded scenarios. 
Additionally, we compare the memory usage and computational complexity of the U-Net variants to our proposed architecture. 
Moreover, we conducted extensive experiments on the Decathlon prostate dataset~\cite{DECA} and the BraTs 2020 dataset~\cite{BRATS1, BRATS2, BRATS3} to prove the effectiveness of the proposed framework in various experimental settings and compare them with state-of-the-art techniques to illustrate the benefits of our approach.

\textbf{The key contributions of this work are:}
\begin{enumerate}
    \item Our architecture network improves the prediction quality on $16 \times 16$ input resolution on the Decathlon dataset by $5.5\%$ and on the BraTS dataset reaches almost the same scores as the baseline using $32 \times 32$ input resolution while increasing the throughput rate only by $0.1$ images per second.
    %\item our architecture network improves the prediction quality of the low-resolution input image by leveraging additional information of higher resolution ground truths without adding significant additional computations.
   \item We can extend any pre-existing U-Net-like architecture to a our architecture to use higher-resolution ground truths efficiently.
   \item We present detailed ablation studies and analysis of the results compared our framework to comparable segmentation methods on the Decathlon prostate dataset and BraTS 2020 to evaluate our method's efficacy.
   %\item This work is open-source and accessible online at \url{https://BlindedLinkForReview}. 
\end{enumerate}

%%%%%%%%%%%%%%%%%%%%%%%%%%%%%%%%%%%%%%%%%%%%%%%%%%%%%%%%%%%%%%%%%%%%%%%%%%%%%%%%%%%%%%%%%%%%%%%%%%%%%%%%%%%%%%%
% 2. SECTION RELATED WORKS
%%%%%%%%%%%%%%%%%%%%%%%%%%%%%%%%%%%%%%%%%%%%%%%%%%%%%%%%%%%%%%%%%%%%%%%%%%%%%%%%%%%%%%%%%%%%%%%%%%%%%%%%%%%%%%%
\section{Related Works}\label{sec:rel}

This section discusses the current state-of-the-art U-Net variations. % and lightweight Networks. 

First, the basic U-Net gained enormous popularity in the biomedical field and thus is usually the starting point for any researcher trying to perform semantic segmentation of medical images. 
Compared to other popular semantic segmentation networks, like ResNet~\cite{RESNET}, U-Net~\cite{UNET} not only consists of an encoder part, but also introduces a decoder part, which up-samples the compressed feature maps. 
%Compared to other popular semantic segmentation networks, like ResNet~\cite{RESNET}, U-Net~\cite{UNET} not only consists of an encoder part, which down-samples the input to extract essential features but also introduces a decoder part, which up-samples the compressed input. 
%Furthermore, each encoder layer uses a skip connection to the corresponding decoder layer of the same level.
\begin{table}[ht]
\caption{Comparison of state-of-the-art U-Net based semantic segmentation for medical imaging.\label{check}}
% \caption{Comparison of state-of-the-art and SILOP on their major source of improvement. Aff. stands for AffinityNet and O.P. of object perimeters.\label{check}}

%\resizebox{\columnwidth}{!}{
\centering 
{%\fontsize{9pt}{9pt}\selectfont
\begin{tabular}{lcccc}
\hline
Methods& \begin{tabular}{@{}c@{}}Efficiency\end{tabular}     & \begin{tabular}{@{}c@{}}Resolution \end{tabular} & \begin{tabular}{@{}c@{}}Loss\end{tabular} & \begin{tabular}{@{}c@{}}Architecture \end{tabular}\\ \hline
ACU-Net~\cite{tan2021multimodal}              & \xmark  &   \xmark &   \xmark  &   \cmark\\
A. Myronenko~\cite{myronenko20193d}         & \xmark  &   \xmark &   \cmark  &   \cmark\\
MH U-Net ~\cite{ahmad2021mh}          & \xmark  &   \xmark &   \xmark  &   \cmark\\
Jiang et al.~\cite{jiang2020two}         & \xmark  &   \cmark &   \cmark  &   \cmark\\
E1D3 U-Net~\cite{bukhari2021e1d3}        & \xmark  &   \xmark &   \xmark  &   \cmark\\
Peiris et al.~\cite{peiris2021reciprocal}      & \xmark  &   \xmark &   \cmark  &   \xmark\\
Zabihollah et al.~\cite{zabihollahy2019automated} & \xmark  &   \xmark &   \xmark  &   \cmark\\
Nie and Shen~\cite{nie2019semantic}        & \xmark  &   \xmark &   \cmark  &   \cmark\\
Zhu et al.~\cite{zhu2019boundary}        & \xmark  &   \xmark &   \cmark  &   \cmark\\
Wang et al.~\cite{wang2019deeply}          & \xmark  &   \xmark &   \cmark  &   \cmark\\
U-Net++~\cite{UNET++}             & \xmark  &   \xmark &   \xmark  &   \cmark\\
U-Net3+~\cite{UNET3+}           & \cmark  &   \xmark &   \cmark  &   \cmark\\
ELU-Net~\cite{ ELUNET}             & \cmark  &   \xmark &   \cmark  &   \cmark\\\hline
Ours                     & \cmark  &   \cmark &   \cmark  &   \cmark \\
\end{tabular}}
\end{table}
  
However, the original U-Net was proposed in 2015, and since then, much progress has been made in deep learning. 
Therefore, several extensions to U-Net were proposed:
ACU-Net~\cite{tan2021multimodal} replaces the convolutional layers of U-Net with deep separable convolutional layers. 
Moreover, Tan et al. utilize residual skip connections and an active contour model. 
In \cite{myronenko20193d}, the authors added a variational auto-encoder branch to a 3D U-Net and achieved first place at the BraTs 2018 challenge. 
MH U-Net~\cite{ahmad2021mh} is a multi-scale hierarchical-based architecture that introduces a hierarchical block between encoder and decoder for acquiring and merging features to extract multi-scale information.  
Jiang et al.~\cite{jiang2020two} proposed a two-stage cascaded U-Net to segment the substructures of brain tumors from coarse to fine; with this method, they achieved first place in the BraTs 2019 challenge. 
E1D3 U-Net~\cite{bukhari2021e1d3} extends the 3D U-Net to use three decoder branches instead of one, where each decoder segments one of the hierarchical regions of interest. 
Peiris et al.~\cite{peiris2021reciprocal} used dual reciprocal adversarial learning approaches. 
The authors followed the Virtual Adversarial Training approach by generating more adversarial examples by adding some noise to the original patient data. 
The resulting network achieved overall better results on BraTs 2021, showing that networks benefit from higher robustness when dealing with multi-modal MRI datasets. 
Zabihollahy et al.~\cite{zabihollahy2019automated} used a separate U-Net per modality on a prostate dataset. 
One was tasked to segment the whole prostate gland, and the other to segment the central gland. 
Nie and Shen~\cite{nie2019semantic} tackled the problem of detecting high-quality contours in prostate images by employing a U-Net with a semantic-guided encoder feature learning strategy and a soft contour constraint mechanism. 
Zhu et al.~\cite{zhu2019boundary} also focused on improved contour segmentation for prostate data. 
They proposed using an S-Net architecture, an encoder-decoder-based network with a boundary-weighted loss function, and additional transfer learning for small datasets. 
Wang et al.~\cite{wang2019deeply} proposed to use deep supervision and group dilated convolution to segment the prostate on magnetic resonance imaging (MRI) with a 3D U-Net. 
\begin{figure*}[ht]
\centering
\includegraphics[width=140mm,scale=0.80]{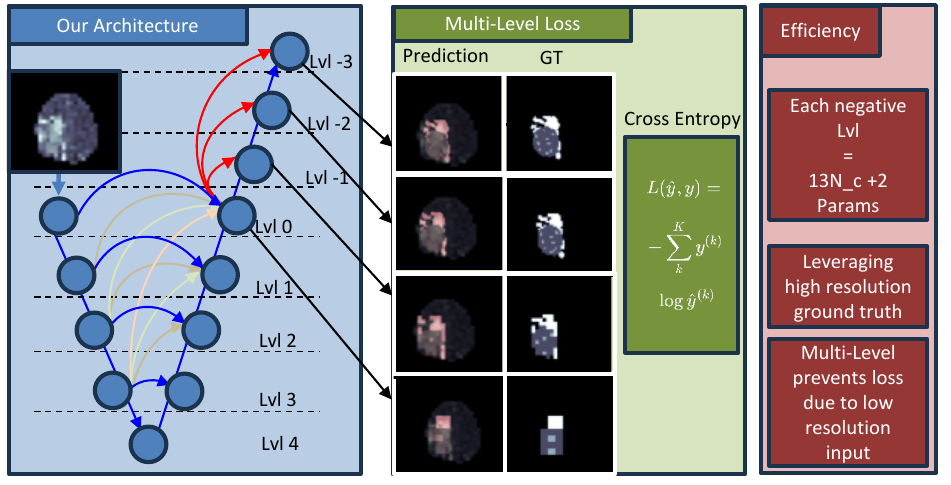} % width=50mm,scale=0.50 width=\linewidth
\caption{Overview of our architecture. 
Our architecture extends the conventional "U" shape of the network to produce higher-resolution outputs. In training, we compare the output at different sizes to the correspondingly reshaped ground truths.\label{fig2}}
\end{figure*}

Another extension is U-Net++~\cite{UNET++}, which replaced the simple skip connection with dense nested connections. 
A simple skip connection concatenates the source layer's output to the target layer's input. 
The dense nested skip connection on U-Net++ added multiple convolutional layers between the source and target layer, and the output of every previous layer will be used in the input of the next layer. 
Thus, U-Net++ achieves higher prediction quality at the cost of much more computational effort. 
U-Net3+~\cite{UNET3+} was proposed to address the high computational costs of U-Net++. 
Instead of dense nested skip connections, it uses full-scale skip connections, which means that every layer in the decoding part of the network also uses the output of the corresponding encoder layer and the downscaled output of encoder layers before the corresponding layer. 
Moreover, each decoder layer has skip connections to all previous decoder layers. 
ELU-Net, a recent U-Net iteration, uses a reversed approach by adding skip connections between decoder layers with the corresponding encoder layers and all subsequent encoder layers. % after up-scaling. 
However, they do not utilize the skip connection between the encoder layers. 
Hence, their network reaches higher prediction quality than the original U-Net with less computation. 
Table.~\ref{check} highlights aspects of improvement in the state-of-the-art for U-Nets.
We observed only one method made use of low resolution inputs and very few concentrated on the cost of deployment.
However, most state-of-the-art approaches rely optimizing the U-Net vase architecture and its loss function.
Moreover, we see that our network is the only approach with a focus on minimizing the cost of deployment by making use of lower resolution inputs.

%%%%%%%%%%%%%%%%%%%%%%%%%%%%%%%%%%%%%%%%%%%%%%%%%%%%%%%%%%%%%%%%%%%%%%%%%%%%%%%%%%%%%%%%%%%%%%%%%%%%%%%%%%%%%%%
% 3. SECTION FRAMEWORK
%%%%%%%%%%%%%%%%%%%%%%%%%%%%%%%%%%%%%%%%%%%%%%%%%%%%%%%%%%%%%%%%%%%%%%%%%%%%%%%%%%%%%%%%%%%%%%%%%%%%%%%%%%%%%%%
\section{Our Framework}

%\begin{multicols}{3}
%\end{multicols}
Fig.~\ref{fig2} presents an overview of our our architecture and our proposed loss function. 
We can use any U-Net as a basis for our architecture since all U-Nets share the standard structure of one encoder and decoder.

First, the U-Net input image runs through the encoder.  
The purpose of the encoder is to extract high-level features, which is achieved by down-sampling the input through each level. 
Next, the decoder takes the output of the encoder. 
Usually, there are as many decoder levels as encoder levels. 
The input for the decoder stems from the last encoder layer, in the case of the first decoder layer, or the previous decoder layer and the encoder layer from the same level, for the subsequent layers.
However, some U-Net variations, also concatenate the feature maps from other decoder layers or encoder layers, with or without additional convolutional layers between them. 
The decoder's purpose is to retrieve spatial information. 
%This is achieved by reversing the down-sampling operation of the encoder of the corresponding level and thus gradually retrieving the original resolution of the input. 
Finally, the last decoder layer generates the semantic segmentation result using a SoftMax function. 
This architecture, consisting of an encoder and decoder, forms the base of one of the most popular networks for biomedical image segmentation. 
  
However, as we already saw in Section~\ref{sec:rel}, the original U-Net has undergone several iterations over the years. 
Nevertheless, to the best of our knowledge, they did not target their performance by reducing the input resolution in an efficient way with regard to the prediction quality.

In application, users often reduce the input resolution if the chosen architecture does not reach the desired performance requirements. 
Nonetheless, the reduction of the input resolution comes with a dramatic loss in prediction quality, although the available data quality would allow
higher quality predictions. 
Moreover, in our experiments, we observed that the bulk of computational load stems from the input resolution rather than the output resolution. 
In other words, reducing (or increasing) the number of decoder layers does not affect the network's overall complexity by a considerable degree. 
Conversely, any change in the number of encoding layers has vast implications. 
Furthermore, higher-resolution ground truths are still available when reducing input resolution for performance reasons. 

Thus, we propose our architecture. 
Our architecture does not interfere with the baseline U-Net in any way but instead adds more up-scaling layers to the decoder part.
This way, we can extend any U-Net-like architecture to our architecture without much effort. 
The additional layers help our architecture to predict a higher resolution than the input to compensate for the lost detail in the initial input compression. 
Our experiments showed that it is most efficient if we utilize simple transposed layers and up-scaling of the input only by a factor of two at the time.  
Therefore, we will need more up-scaling layers if we significantly reduce the input resolution. 
However, using fewer transposed layers with a higher up-scaling factor will reduce the prediction quality significantly. 
%Furthermore, we decided not to utilize skip connections between the new layers and the base network, as those would lead to a network of similar complexity to using the high input resolution to begin with. 
We will compute the number of additional parameters here to prove that the proposed addition of up-scaling layers does not increase the computational load significantly. 
The up-scaling layer consists of a ConvTranspose2d and a Conv2d, which we define as $C^T$ and $C$, respectively. 
The trainable weights $W_t(\cdot)$ of both layers can then be described as:
\begin{equation*}
    \begin{split}
        &W_t(C) = W_t(C^T)= \\
        & (H_{kernel} \times W_{kernel} \times c_{in} + 1) \times c_{out},        
    \end{split}
\end{equation*}

where $c_{in}$ is the number of channels if the layer input, $c_{out}$ is the number of channels after the convolutional layer, $H_{kernel}$ and $W_{kernel}$ are the dimensions of the used convolutional kernel. 
In our architecture $c_{in}$ and $c_{out}$ are equal to the number of classes $N_c$, and for the ConvTranspose2d layer we use a $2 \times 2$ kernel, and for the Conv2d layer a $3 \times 3$ kernel. 
That means each up-sampling operation will add the following number of parameters: 
$$\#Params = (4N_c  +1) \times N_c + (9N_c  +1) \times N_c = 13N_c^2+ 2 N_c.$$ 
However, we perform the up-scaling layer multiple times until we reach the desired output resolution. 
For example, suppose we want to train with a compact resolution of $16 \times 16$, and we have a ground truth available of at least $256 \times 256$ resolution. 
In that case, we must perform four up-sampling operations in our architecture, doubling the output resolution each time. 
The number of classes we used for the Decathlon prostate dataset is $N_c=1$, adding up to an additional $60$ trainable parameters. 
Considering that the baseline models consist of several million trainable parameters, those $60$ are negligible for additional computing efforts. 
However, regarding memory consumption, we must pay a higher price to store and load higher-resolution outputs and ground truths. We will provide an in-depth analysis of additional memory costs in Section~\ref{sec:experiments}. 
  
Furthermore, to take more advantage of our architecture, we propose a new loss function: 
  
$$L_{sum} = \sum_{i=0}^{m} L_i(F_i(Y),\hat{Y}_i)$$ 
  
Where $L_i$ denotes our loss of choice, for example, Cross Entropy, $Y$ is the input image, $\hat{Y}_i$ is the output of the $i_{th}$ transposed layer, where $\hat{Y}_0$ denotes the prediction with the input resolution, $m$ is the number of up-scaling layers we need, to reach the desired output resolution defined as $M$.% and with each increment of $i$ the resolution doubles. 
We observed the best results when choosing $M$ as high as possible and therefore strive to have at least one term compared to the most detailed ground truth to our availability. 
Incorporating all losses for each up-scaling stage does not add much complexity since the Cross-Entropy Loss is relatively simple, and extracting the output of each up-scaling layer comes at no cost. 
%####################################################### 
%####################################################### 

We also experimented with other methods of reducing the resolution. 
One apparent approach was to use artificially up-scaled ground truth, which would allow us to boost the prediction quality without adding many complex operations to the network architecture.
However, simply up-scaling the ground truth resolution did not lead to improved prediction quality, as fine-grained details are not added when up-scaling the ground truth naively. 
%Using high-resolution input images with low-resolution ground truths has the opposite effect: lower prediction quality and higher computational complexity. 
Furthermore, extending our architecture with only one up-sampling operation that increases the input resolution by a factor of $16$ instead of four up-sampling operations with a factor of $2$ did lead to predictions, which are similar to the quality of the predictions generated without extra up-sampling. 
We also tried adding skip connections from the encoder to the new up-scaling layers, as ELU-Net and U-Net3+ showed that those skip connections can significantly improve prediction quality while simultaneously being cost-efficient when compared to dense connections. 
However, we observed that adding such skip connections did not contribute to better model predictions but added almost as many parameters as using higher input resolution from the start. 
our architectures up-scaling layers contribute very few parameters to the model because they take as input feature maps with as many channels as classes.
 However, adding skip connections from other layers in the base network would increase the number of parameters by a lot, because inside the base network, the number of channels is balanced by a reduction of feature map resolution.
 But by combining those layers with our architectures up-scaling we would have to increase their resolution again, generating a huge number of additional operations.
%Feature maps of low levels have a high resolution but few channels. 
%As we increase the level of our network, we gain more channels but with lower resolution. 
%Suppose we connect higher-level feature maps to J-Net up-scaling layers. 
%In that case, we have to re-scale their resolution accordingly, which leads to a high-resolution feature map with many channels as input for our up-scaling layer. 
%Therefore, we would end up with huge feature maps in our up-scaling layer, which would require more computations to the network than almost any layer in the base network. 
%###################################################
%Furthermore, we also experimented by adding a classification loss.
%However, adding a classification did not contribute to more precise predictions, while additional computations were added due to the new fully connected layers.
%Our intuition for this decision was inspired by weakly supervised semantic segmentation.
% Weakly supervised networks predict pixel-wise predictions by being only trained with classification labels.
%This shows that the classification loss already gives the model beneficial feedback, even for pixel-wise prediction tasks.
%As for classification loss, we used Cross Entropy, and for the predictions, we retrieved the output from the last encoder layer and applied a flattening operation and a fully connected layer to obtain an output in classification form, meaning a one-hot encoded vector with as many dimensions as classes.
%###################################################
On the other side, we also experimented with adding skip connections between the additional up-scaling operations, as those would make it easier for the gradient to traverse the now deeper network. 
In this case, we would only add as many channels as classes to the up-scaling layers. 
Compared to connections to the base network, the additional computations are acceptable and still much less than doubling the input resolution. 
For our example of a $16 \times 16$ input image and one class, we will end up with the following number of parameters: % $ \#Params $: 

    % $$\#Params = 2(2 (1)^2 \times 4 + 2(1)^2 \times 9 +2) = 4(17(1)^2 + 2)=76,$$ 
\begin{equation*}
    \begin{split}
        \#Params &= [13 \times 1^2 + 2 \times 1^2]2\\
        &+[4 \times 1^2 +1]7+ [9 \times 1^2 +1]2\\
        &+[(4 \times 2 +1) + (9 \times 2 +1)]2\\
        &=159,
    \end{split}
\end{equation*}

\begin{figure}[t]
\centering
\includegraphics[width=0.8\columnwidth]{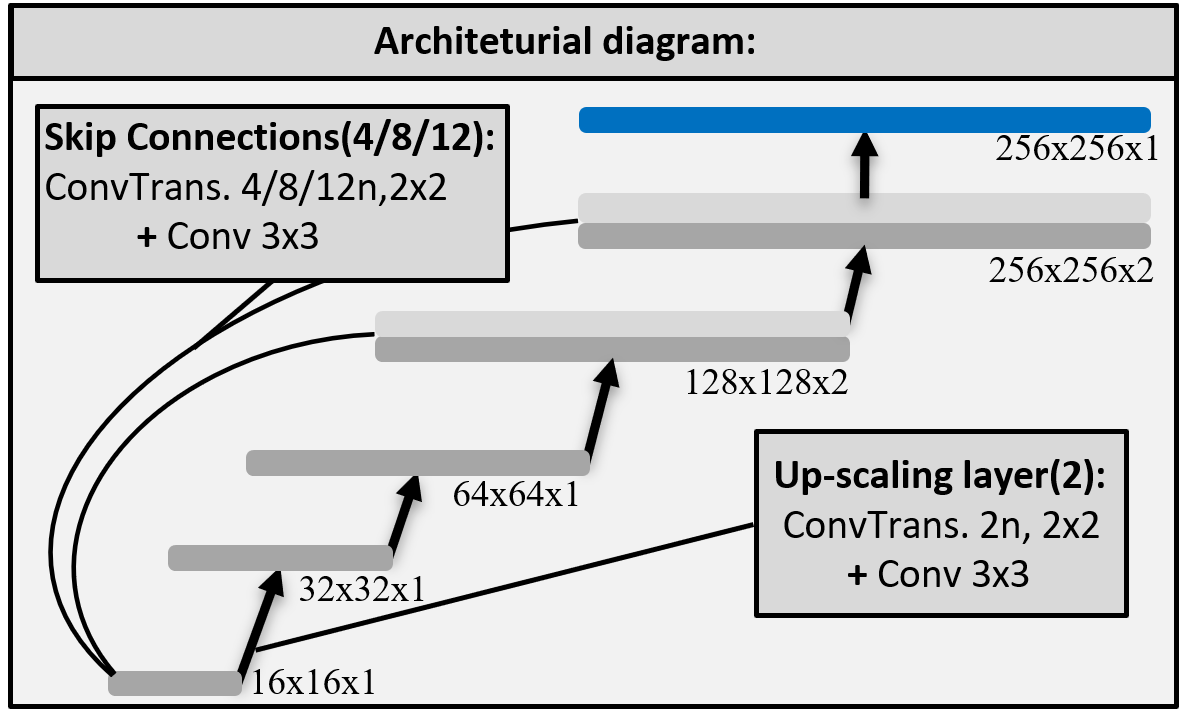} % width=50mm,scale=0.50 width=\linewidth
\caption{Detailed illustration of our architecture up-scaling layers with skip connections for the case of a $16 \times 16$ input image with one class.
\label{arch}}
\end{figure}

where the first term refers to normal two up-scaling layers, and the second term describes the stretching of the intermediate result for the skip connections, where we performed transposed convolution with a higher up-scaling factor each followed by one normal convolution, to reach the desired resolution.
The higher up-scaling factor translates to performing the ConvTranspose2D operation multiple times, in our case $3$ and $times$.
The last term describes the two up-scaling layers, that used the additional skip connections, thus the input channels increase from one to two.
Compared to the $60$ parameters without skip connections within our architecture extension. 
Unfortunately, we also increase memory consumption, which we must consider when deciding whether to use these skip connections. 
These additional skip connections improve the prediction quality, and thus, it depends on the individual use case whether we have enough hardware capacity left to utilize them.
Fig.~\ref{arch} illustrates our architectures up-scaling layers with skip connections in detail.
Note that we did not use the skip connection for the second up-scaling layer as this showed to be not as effective.
%###################################################
%When observing the used decathlon and BraTS datasets, we noticed that every image has only one target object.
%In the case of the decathlon dataset, the target is always located at a similar location and size.
%Hence, we also experimented with cropping the input image to achieve a smaller input resolution while simultaneously preventing any loss in detail or accidentally removing parts of the target object.
%We experimented with two versions.
%The first was the real-world scenario where we used a first network to predict the area of interest to decide the crop location, with an apriori-determined crop size. 
%The first network used the low-resolution input image to reduce the computation time but still managed to capture the target object entirely almost all the time. 
%The second version used ground truth segmentation and created crops of the target object that filled the image, thus minimizing the amount of background pixels.
%In both versions, we trained a second network to predict the crops.
%However, both methods failed to generate predictions at the same level as with the original image.
%%%%%%%%%%%%%%%%%%%%%%%%%%%%%%%%%%%%%%%%%%%%%%%%%%%%%%%%%%%%%%%%%%%%%%%%%%%%%%%%%%%%%%%%%%%%%%%%%%%%%%%%%%%%%%%
% 4. SECTION RESULTS
%%%%%%%%%%%%%%%%%%%%%%%%%%%%%%%%%%%%%%%%%%%%%%%%%%%%%%%%%%%%%%%%%%%%%%%%%%%%%%%%%%%%%%%%%%%%%%%%%%%%%%%%%%%%%%%
\section{Results and Discussion}\label{sec:experiments}

This section will review the experimental setup used, the results obtained, and the ablations and insights we draw from them. 
  
We start with an overview of the used hardware, the software libraries, and their respective versions.

We trained the networks on a CentOS 7.9 Operating System executing on an Intel Core i7-8700 CPU with 16GB RAM and 2 Nvidia GeForce GTX 1080 Ti GPUs until saturation. 
%##UPDATE PLS########################################## 
%We executed our scripts with the following software versions: %CUDA 11.5, Pytorch 3.7.4.3, and Torchvision 0.11.1. 
%##UPDATE PLS########################################## 
For our performance evaluation, we also ran the networks on Nvidia's Jetson Nano, a small edge device for deep learning deployment. 
We used ELU-Net~\cite{ELUNET} as the backbone for our architecture. 
  
We used the Jaccard Coefficient and the Dice Coefficient as the evaluation metrics for all experiments. 
The Jaccard is defined as follows: 
{%\fontsize{9pt}{9pt}\selectfont
\begin{equation}
\centering
Jaccard = \frac{1}{N} \sum_{i=1}^N \frac{ p_{i,i}}{\sum_{j=1}^N p_{i,j} + \sum_{j=1}^N p_{j,i} - p_{i,i}  }
\end{equation}
}
The Dice score is defined as:
{%\fontsize{9pt}{9pt}\selectfont
\begin{equation}
\centering
DICE = \frac{2}{N} \sum_{i=1}^N \frac{ p_{i,i}}{\sum_{j=1}^N p_{i,j} + \sum_{j=1}^N p_{j,i}  }
\end{equation}
}
\noindent where $N$ is the total number of classes, $p_{i,i}$ the number of pixels classified as class $i$ when labelled as class $i$. $p_{i,j}$ and $p_{j,i}$ are the number of pixels classified as class $i$ that were labelled as class $j$ and vice-versa, respectively. 
Those two metrics are the gold standard for medical semantic segmentation. 
The Jaccard coefficient measures the ratio between the number of correctly predicted pixels and the overall number of pixels in the prediction and the ground truth. 
In other words, the degree of overlap between prediction and ground truth. 
Meanwhile, the Dice coefficient measures the ratio between the number of correct predicted pixels and the disjoint union of the prediction and the ground truth, a coefficient between the similarity of prediction and ground truth. 
  
We evaluate our method on two medical datasets the Decathlon prostate dataset, and the Brain Tumor Segmentation (BraTS) Challenge 2020 dataset.

\textbf{Decathlon:}
\begin{table}[t]
%\resizebox{\columnwidth}{!}{
\centering 
{%\fontsize{8pt}{8pt}\selectfont
\caption{Comparison of U-Net, ELU-Net, and our architecture on the Decathlon dataset at different input resolutions.\label{tabdeca}}
\begin{tabular}{lccc}
\hline
Method         & Resolution       & Dice  & Jaccard \\ \hline
U-Net~\cite{UNET} &         &  98.8 & 97.7 \\
ELU-Net~\cite{ELUNET}& 320x320           & 98.4 & 96.9 \\
Ours       &         & - & - \\ \hline
U-Net &         &  98.8 & 97.6 \\
ELU-Net& 160x160           & 98.7 & 97.4 \\
Ours       &         & 98.7 & 97.4 \\ \hline
U-Net&             &  98.4 & 96.9 \\
ELU-Net&   80x80           & 98.7 & 97.4 \\
Ours &           & \textbf{98.7} & \textbf{97.4} \\ \hline
U-Net &          &  96.0 & 92.4 \\
ELU-Net&    32x32      & 96.5 & 93.3 \\
Ours       &    & \textbf{98.0} & \textbf{96.1} \\ \hline
U-Net&         &  93.2 & 87.3 \\
ELU-Net  &   16x16      & 93.0 &  86.9 \\
Ours  &            & \textbf{96.2} & \textbf{92.6} \\ \hline
\end{tabular}}

\end{table}
The Decathlon segmentation challenge is a multi-modal prostate dataset with the task of localizing and detecting the prostate peripheral and transition zones. 
The dataset includes $32$ volumetric scans.
Each scan is available in two modalities for each scan: the MRI-T2 and the Apparent Diffusion Coefficient (ADC). 
Furthermore, each scan also includes corresponding segmentation masks. 
We separated the scans into 2D slices, resulting in $1806$ slices overall.
We split the dataset into train and validation sets in a $2/3$ to $1/3$ ratio. 
In our experiments, we observed that adding the ADC modality to the model input had little to no effect on the model's prediction quality. 
Therefore, we only used the MRI-T2 maps for training, which further reduces computational complexity. 
Moreover, we combined the prostate peripheral zone and the transition zone into a single `positive' class as the transition zone is barely present in the dataset, and the models could not detect this class. 
Most 2D slices were available in $320 \times 320$ resolution. 
We stretched the image to this size using OpenCV's inter-linear interpolation scheme in cases where the provided resolution was lower than $320 \times 320$. 
  
In Table~\ref{tabdeca}, we compare the results on the Decathlon datasets. 
We compared a basic U-Net, ELU-Net, and our architecture. 
We list the Jaccard and Dice results on the validation set and perform the experiments with different input resolutions.
For our architecture, we always used the $320 \times 320$ resolution ground truth, except for input resolution $32 \times 32$ and $16 \times 16$, for which we used a $258 \times 258$ ground truth resolution. 

ELU-Net achieves a $0.8\%$ Jaccard improvement over U-Net on the highest input resolution. 
%We will use the ELU-Net scores as our upper bound for the $320 \times 320$. 
However, training the ELU-Net with a slightly lower resolution seems to affect the overall prediction quality positively. 
Note that when the model takes a lower input resolution, we will stretch the predictions using OpenCV's inter-linear interpolation to the resolution of the original ground truth and use only the scores compared to the authentic ground truth. 
Nevertheless, if we further reduce the input resolution, the prediction quality of ELU-Net (and U-Net) diminishes, as expected. 
We observe that our architecture can maintain higher prediction quality on lower input resolutions than the other networks.
Furthermore, we notice that the more we reduce input resolution, the more significant the difference between our architecture and ELU-Net. 
The difference between ELU-Net at $80 \times 80$ input resolution is non-existent, whereas, with $32 \times 32$ input resolution, the difference is almost $3\%$ Jaccard score. 
At $16 \times 16$ input resolution, our architecture and ELU-Net are separated by $5.5\%$ Jaccard score. 
We can explain this because ground truth information diminishes greatly by reducing its resolution, while our architecture keeps using the high resolution and thus high information ground truths. 
Therefore, our architecture is not as much affected by the lower input resolution.

\begin{table*}[ht]
\caption{Comparison of U-Net, ELU-Net, and our architecture on the BraTs dataset at different input resolutions.\label{tabbrats}}

%\resizebox{\columnwidth}{!}{
\centering 
{%\fontsize{9pt}{9pt}\selectfont
\begin{tabular}{lccccccc}
\hline
Method & Input                  & ED Dice & ED Jaccard  & NCR Dice & NCR Jaccard  & ET Dice & ET Jaccard \\ \hline
U-Net~\cite{UNET}&              & 86.9    & 82.1        &    88.3 &      85.6   &    89.2 &      85.6   \\
ELU-Net~\cite{ELUNET} & 256x256   & 88.3    & 83.5        &   88.5  &      85.9   &    90.0 &      86.4    \\
Ours          &         & -    & -        &    -   &      -   &    - &      -     \\ \hline
U-Net &             & 85.4   &  79.9     &  87.4 &    84.5   &   87.6 &      83.7   \\
ELU-Net &128x128     &  85.4   & 80.0        & 86.9    &  84.3       &   87.7  &  83.8        \\
Ours &                  &   \textbf{87.6}  &  \textbf{83.5}       & \textbf{88.2}    &  \textbf{85.9}       &  \textbf{89.2}   &   \textbf{86.2}      \\ \hline
U-Net &             &   80.6  &  74.7       &  84.7   & 82.1      &  84.0   &   80.0       \\
ELU-Net&64x64       &    81.4 & 75.4        & 85.3    &   82.6      &  84.8   & 80.8         \\
Ours &                  &    \textbf{85.4} & \textbf{80.9}        & \textbf{86.6}    &   \textbf{84.4}      &  \textbf{87.5}  & \textbf{84.2}         \\ \hline
U-Net &             &  74.7   &   68.9     &  82.1   &    79.6     &  80.0   &    76.4      \\
ELU-Net&32x32       &  75.2   &  69.3       &  82.7   &    80.3     & 80.7    &   77.2       \\
Ours &                  &  \textbf{82.2}   &    \textbf{77.4}     &  \textbf{84.7}   &  \textbf{82.5}       &  \textbf{84.7}   &      \textbf{81.3}    \\ \hline
U-Net&             &  67.3   &   62.6      &  78.8   &    77.0     &  75.5   &    73.0      \\
ELU-Net&16x16      &  67.5   &  62.8       &  79.5   &    77.7     &  76.8   &   74.0       \\
Ours &                  &   \textbf{75.5}  &     \textbf{70.7}    &   \textbf{81.6}  &     \textbf{79.8}    &  \textbf{78.7}   &      \textbf{76.0}    \\ \hline
\end{tabular}}
\end{table*}

\textbf{BraTs:} The BraTS dataset is a multi-modal brain tumor segmentation in Magnetic Resonance Images. 
The dataset includes 369 multi-modal scans with their corresponding expert segmentation masks. 
The available modalities are T1, T1c, T2, and T2 Fluid Attenuated Inversion Recovery (FLAIR). 
The dataset contains annotations for the GD-enhancing tumor (ET), the peritumoral edema (ED), and the necrotic and non-enhancing tumor core (NCR). 
We separated each scan into its set of slides, resulting in 50,764 slides, and cropped them to the size $256 \times 256$. 
We used a $85\%$ to $15\%$ train validation split for our experiments. 
  
In Table~\ref{tabbrats}, we compare the results of a basic U-Net, ELU-Net, and our architecture on the BraTs datasets. 
We used $256 \times 256$ resolution ground truth for all our input resolutions. 
As typical in this dataset, we evaluate the results for each class separately. 
This time, our architecture outperforms ELU-Net with the same input resolution and can achieve better results than ELU-Net with the next higher input resolution for the ED class and almost the same for the other two classes. 
For the input resolution of $128 \times 128$, our architecture improves on ELU-Net by $2.5\%$ Jaccard on average and loses to ELU-Net with the input of $256 \times 256$ resolution $0.07\%$ Jaccard on average. 
Similar to the evaluation on Decathlon, our architecturet shows the most significant improvement with input resolution $16 \times 16$, achieving an average $4\%$ better Jaccard prediction result. 
Note that when evaluating before saturation, our architecture has an even higher lead than ELU-Net, showing that the up-scaling layers and high-resolution ground truths lead to fast learning. 

%\textbf{Model Complexity:} 
%\begin{table}[ht]
%%\resizebox{\columnwidth}{!}{
%\centering 
%{%\fontsize{9pt}{9pt}\selectfont
%\caption{Number of model parameters for the tested models.\label{tabparam}}
%\begin{tabular}{lc}
%\hline
%Method         & $\#$Parameters (M.) \\ \hline
%U-Net~\cite{UNET}& 23.38 \\
%ELU-Net~\cite{ELUNET}& \textbf{3.38} \\ 
%Ours       & \textbf{3.38}\\ \hline
%\end{tabular}}

%\end{table}

%Here, we will evaluate our model's hardware requirements compared to the baseline models' %requirements, as this was the central motivation of the work. 

Here, we will evaluate our model's hardware requirements compared to the baseline models' requirements, as this was the central motivation of the work. 
%Table~\ref{tabparam} lists the number of parameters of a classic U-Net, ELU-Net, and J-Net. 
%We can see that the added layers of J-Net contribute only a few extra parameters, especially compared to the classical U-Net. 

%First, we will analyze the number of parameters for the compared architectures:
%The classical U-Net has $23.38$ M. parameters, ELU-Net $3,377,201$, and J-Net $3,377,380$.
%We can see that the added layers of J-Net contribute only very few extra parameters, especially compared to the classical U-Net. 

\begin{table}[ht]
%\resizebox{\columnwidth}{!}{
\centering 
{%\fontsize{8pt}{8pt}\selectfont
\caption{Complexity and memory consumption of the tested models with respect to the input resolution of one image with one channel.\label{tabcomp}}
\begin{tabular}{lccc}
\hline
Method         & Resolution& Complexity (GMac)& Memory (mb)\\ \hline
U-Net~\cite{UNET}  &         & 71.16 & 1191.41 \\
ELU-Net~\cite{ELUNET}& 320x320 & 29.98 & 868.36  \\
Ours       &         & -     & -       \\ \hline
U-Net&         & 17.79 & 297.85  \\
ELU-Net& 160x160 &  \textbf{7.5}  & \textbf{217.09}  \\
Ours       &         &  \textbf{7.5}  & 219.43  \\ \hline
U-Net &         & 4.45  & 74.46   \\
ELU-Net&   80x80 & \textbf{1.87}  & \textbf{54.27}   \\
Ours       &         & 1.88  & 58.57  \\ \hline
U-Net &         & 0.71  & 11.91   \\
ELU-Net&   32x32 & \textbf{0.30}  & \textbf{8.68}    \\
Ours       &         & \textbf{0.30}  & 13.68  \\ \hline
U-Net  &         & 0.18  & 2.98    \\
ELU-Net&   16x16 & \textbf{0.07}  & \textbf{2.17}    \\
Ours       &         & 0.08  &  5.5  \\ \hline
\end{tabular}}

\end{table}

%\begin{table}[ht]
%%\resizebox{\columnwidth}{!}{
%\centering 
%{%\fontsize{9pt}{9pt}\selectfont
%\caption{[UPDATE REQUIRED]Complexity, memory consumption, and performance on Jetson Nano of the tested models with respect to the input %resolution, for inference of one image with one channel.\label{tabcomp}}
%\begin{tabular}{lcccc}
%\hline
%Method         & Resolution& Complexity (GMac)& Memory (mb)&Throughput (s/Image)\\ \hline
%U-Net~\cite{Unet}  &         & 71.16 & 1191.41& - \\
%ELU-Net~\cite{Unet}& 320x320 & 29.98 & 868.36 & - \\
%Ours       &         & -     & -      & - \\ \hline
%U-Net~\cite{Unet}  &         & 17.79 & 297.85 & - \\
%ELU-Net~\cite{Unet}& 160x160 &  7.5  & 217.09 &  \\
%Ours       &         &  7.5  & 219.43 &  \\ \hline
%U-Net~\cite{Unet}  &         & 4.45  & 74.46  & - \\
%ELU-Net~\cite{Unet}&   80x80 & 1.87  & 54.27  &  \\
%Ours       &         & 1.88  & 58.57  &  \\ \hline
%U-Net~\cite{Unet}  &         & 0.71  & 11.91  & - \\
%ELU-Net~\cite{Unet}&   32x32 & 0.30  & 8.68   &  \\
%Ours       &         & 0.30  & 13.68  &  \\ \hline
%U-Net~\cite{Unet}  &         & 0.18  & 2.98   & - \\
%ELU-Net~\cite{Unet}&   16x26 & 0.07  & 2.17   &   \\
%Ours       &         & 0.08  &  5.5  &  \\ \hline
%\end{tabular}}

%\end{table}
In Table~\ref{tabcomp}, we compare the complexity, memory requirements, and the time needed to evaluate one image with one channel on Jetson Nano for the different networks, as this was the setting we used with the Decathlon dataset. 
Comparing first the complexity in the number of the Giga Multiply and Accumulate (GMac) operations, we notice that our architecture adds around $0.01$ GMacs to ELU-Net. 
The additional $0.01$ GMacs add relative $14\%$ more complexity between ELU-Net and our architecture at $16 \times 16$ input resolution.
However, at $160 \times 160$, those additional GMacs are negligible relative to the $7.5$ GMacs of ELU-Net. 
As we increase the input resolution size, the relative difference in memory consumption between our architecture and the other networks decreases;
for $160 \times 160$ inputs, our architecture uses less than $1\%$ more memory than ELU-Net and considerably less than U-Net. 
When we compare our architecture's memory requirements to the other networks of the next higher input resolution class, we notice that even for the $16 \times 16$ inputs, our architecture uses $40\%$ less memory than ELU-Net with $32 \times 32$ inputs resolution but reaches on decathlon a prediction quality closer to the results of ELU-Net with $32 \times 32$ inputs than with $16 \times 16$, which is in our eyes an acceptable trade-off. 
%Turning now to the memory requirements, our architecture does not compare as favorably to the other networks. 
For $16 \times 16$ input resolution, our architecture needs more than double the memory than ELU-Net and almost double the U-Net. 
This is mainly because, in our architecture, we need to store the full resolution prediction, while the other networks only deal with predictions and ground truths that are $20$ times smaller. 
\begin{figure}[ht]
\centering
\includegraphics[width=0.7\linewidth]{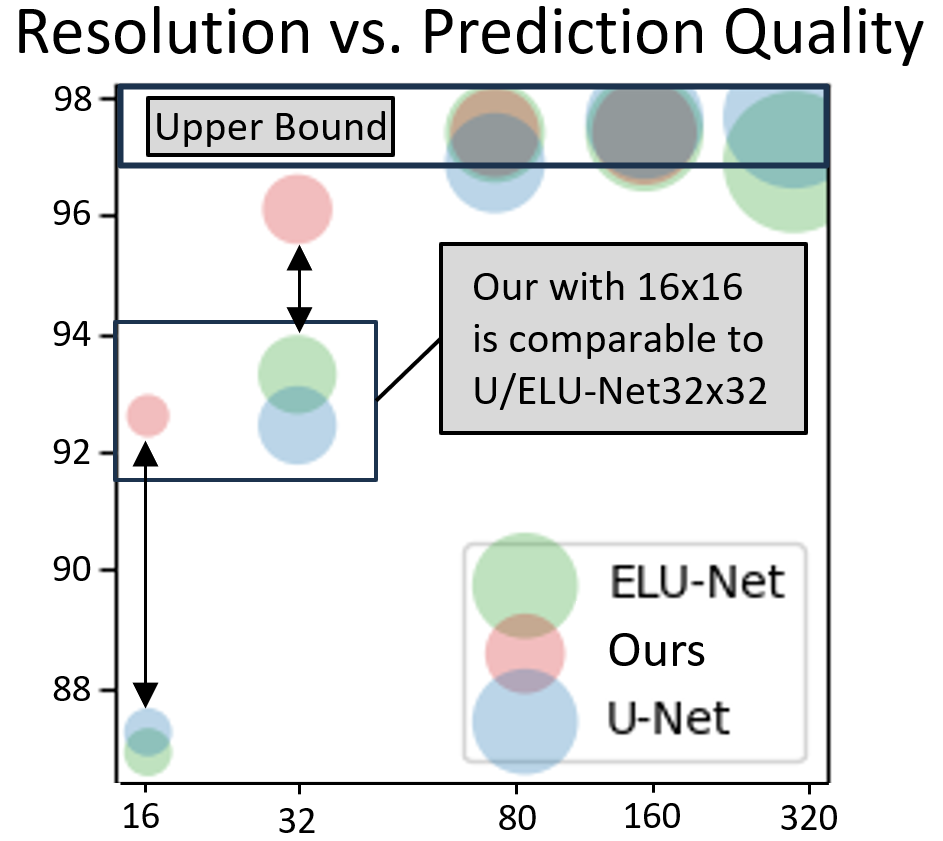} % width=40mm,scale=0.30 width=\linewidth [width=150mm,scale=0.50]
\caption{A comparison between input resolution and prediction quality in Jaccard of the different networks on Decathlon. \label{figscatter}}
\end{figure}
Fig.~\ref{figscatter} visualizes the trade-off between input resolution and prediction quality of the tested U-Nets.
Starting with the smallest input resolution of $16 \times 16$, we notice a considerable gap between our architecture and ELU-Net.
At the same time, our architecture is comparable to ELU-Net, with double the input resolution regarding prediction quality.
The gap between our architecture and ELU-Net, with both using the input resolution $32 \times 32$, shrinks but is still significant.
However, the difference between our architecture and ELU-Net is not noticeable for higher input resolutions.
On the one hand, we can explain that the difference in architecture between the two networks is minimal for higher input resolutions.
On the other side, $97\%$ seems to be the upper boundary for our set-up, and neither the different architectures nor increasing the input resolution seem to change this.
%but is still closer to the ELU-Net with the same input resolution than with the next higher one.
%, but since the last layers are relatively few channels compared to layers inside the backbone network.
\begin{figure}[ht]
\centering
\includegraphics[width=0.9\linewidth]{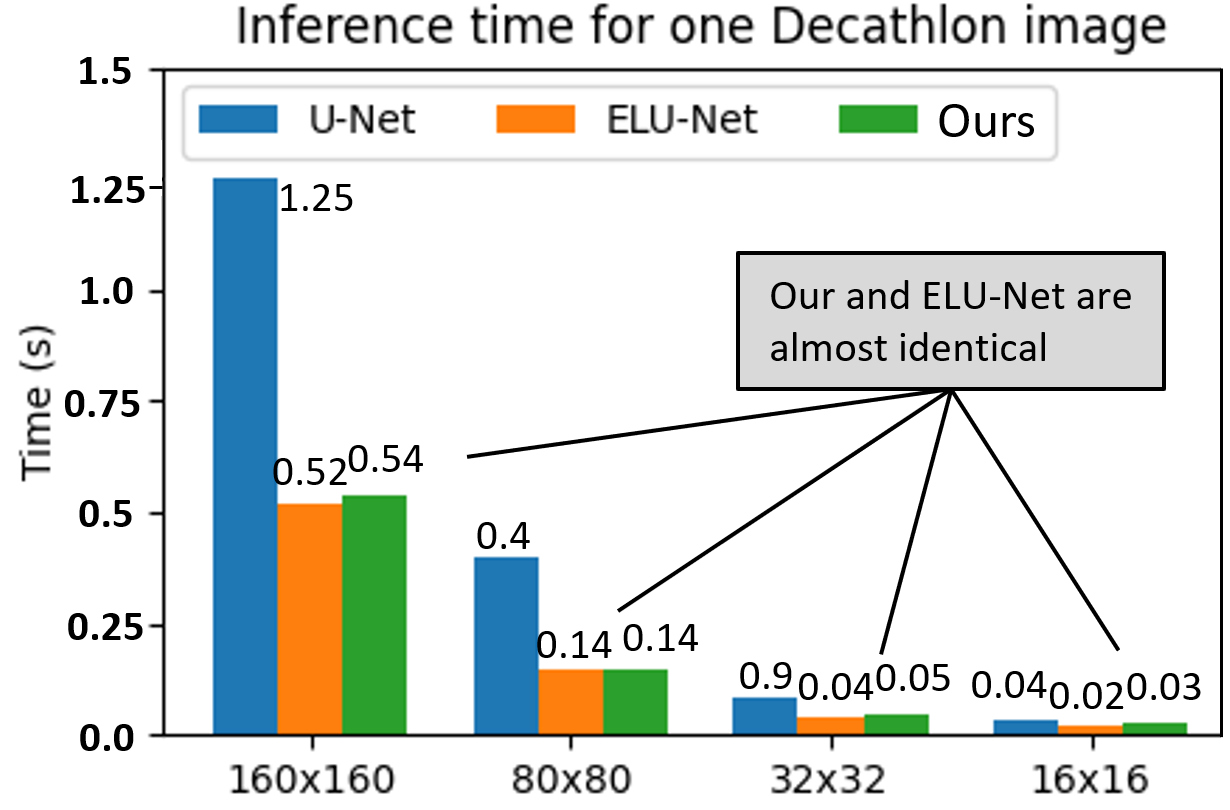} % width=40mm,scale=0.30 width=\linewidth [width=150mm,scale=0.50]
\caption{Inference time in seconds for evaluating one decathlon image in Nvidia's Jetson Nano. \label{figtimes}}
\end{figure}

Fig.~\ref{figtimes} shows the number of seconds it takes to pass one Decathlon image with one channel on Nvidia's Jetson Nano. 
In deployment on Nvidia's Jetson Nano, it becomes clear why we must reduce the input resolution. 
%Training on the already reduced resolution of $80 \times 80$ with the single channel decathlon image takes over two seconds, translating to a total training time of 15 hours on this dataset. 
%Conversely, training on an image of size $16 \times 16$ will take less than 30 minutes. 
%If it comes to the more realistic use case of only deploying an already trained model, notice that
With input resolution $80 \times 80$, we achieve a throughput of one
 image in $0.4$ seconds for the U-Net, translating to $2.5$ images every second.
The more efficient ELU-Net and our architecture need around $0.15$ seconds per image, which is almost seven images per second.
A desirable throughput rate would be closer to 30 images per second; in this case, a live evaluation of a video stream would be possible. 
Rescaling the input to $16 \times 16$ would yield us $0.02$ seconds per image for ELU-Net, translating to $50$ images per second, which is better than our goal.
The additional complexity of our architecture reduces the throughput rate to $0.25$ seconds per image or $40$ images per second, which is still above our target of $30$ images per second.
However, the prediction quality of our architecture with $16 \times 16$ input resolution is comparable to ELU-Net with $32 \times 32$ input resolution.
In this case, ELU-Net has a throughput rate of $0.04$ seconds per image or $25$ images per second, which is over $40\%$ slower than our architecture.
%Now that we have established the importance of reducing the input resolution to achieve an acceptable throughput number, we will compare the performance of J-Net to ELU-Net since we did add some complexity to the model to compensate for the loss in prediction quality due to reduced input resolution. 
%Fortunately, J-Net's reached frames per second are almost identical to ELU-Net, which was our primary goal. 
%We only add $0.005$ seconds on average due toour architecture extensions, which translate to output $0.05$ images less per second for a several percent prediction quality boost. 

\begin{figure*}[t]
\centering
\includegraphics[width=140mm,scale=0.80]{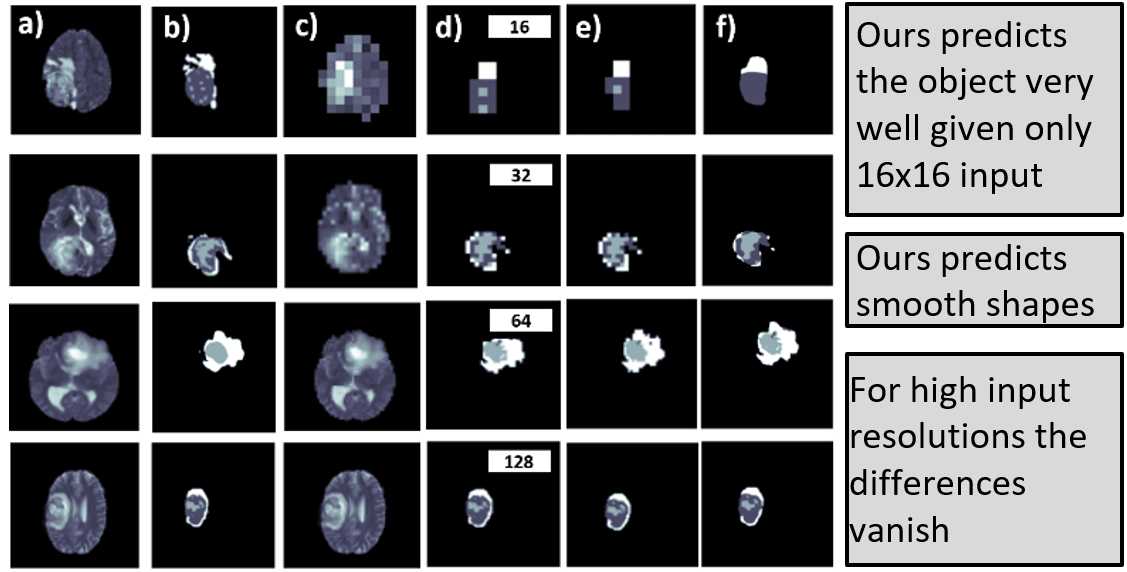} % width=50mm,scale=0.50 width=\linewidth
\caption{Example results of our architecture and ELU-Net for comparison. The a) column shows the T2 brain MRIs in their original resolution. The b) column shows the ground truth in the original resolution and the c) and d) columns the T2 brain MRIs and ground truths in input resolution respectively, the resolution is specified in the top left corner in column d), and the e) column shows the prediction of ELU-Net), and the f) column shows the prediction of our architecture.
\label{exam_brats}}
\end{figure*}

\begin{figure*}[t]
\centering
\includegraphics[width=140mm,scale=0.80]{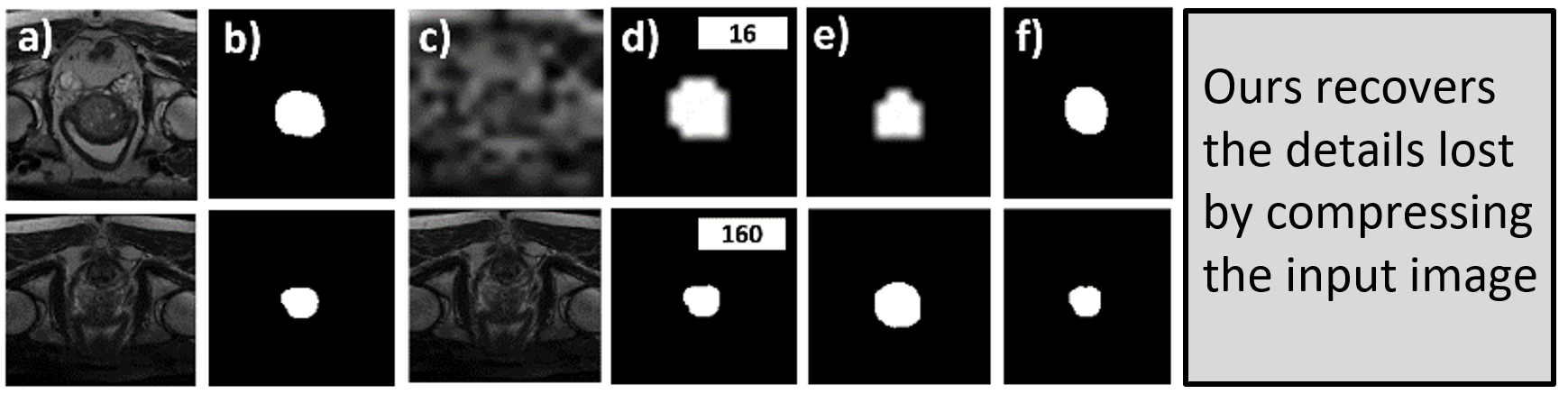} % width=50mm,scale=0.50 width=\linewidth
\caption{Example results of our architecture and ELU-Net for comparison. The a) column shows the T2 brain MRIs in their original resolution. The b) column shows the ground truth in the original resolution and the c) and d) columns the T2 brain MRIs and ground truths in input resolution respectively, the resolution is specified in the top left corner in column d), and the e) column shows the prediction of ELU-Net), and the f) column shows the prediction of our architecture.
\label{exam_deca}}
\end{figure*}

Fig.~\ref{exam_brats} and Fig.~\ref{exam_deca} give a qualitative overview of our architectures results compared to ELU-Net with $16 \times 16$ input.  
Each row of the figures depicts an example image at a different input resolution from BraTs and Decathlon, respectively. 
Column `a)’ shows the MRI-T2 modality, and column `b)’ indicates the corresponding ground truth in the original resolution.
We chose the T2 modality as untrained eyes can spot cancerous regions relatively easily. 
Columns `c)’ and `d)’ show the MRI-T2 modality and ground truth in reduced resolution as indicated in each ground truth.
Column `e)’ and column `f)’ depicts the prediction of ELU-Net and our architecture given the low-resolution input image in column `a).’
Note that ELU-Net was trained only using ground truths of the same resolution as the input, while our architecture was trained using original resolution ground truths.
Fig.~\ref{exam_brats} elucidates how the difference between ELU-Net and our architecture is most noticeable at the lowest input resolution. 
our architectures' low resolution predictions are very detailed, which shows that the network can extract significant amounts of useful information using highly compressed images. 
Similarly, in Fig.~\ref{exam_deca}, even with the $16 \times 16$ input image, our architecture predicts a shape quite like the high-resolution ground truth.
We can summarize that networks can extract much more information from their input than their output suggests and are limited by their output resolution.
Moreover, we observe that our architecture allows us to overcome this limitation.
%\textbf{Ablations:}

%Table~\ref{tab4} compares the difference in prediction quality, parameter numbers, and Jetson Nano performance when using skip connections to not using them in J-Net.
%As we can read in the table, not using the skip connections leads to a middle way between J-Net with connections and ELU-Net, we gain negligible better performance numbers, for $1\%$ o $2\%$ loss in prediction quality.

\section{Conclusion}
In this paper, we have proposed our our architecture, which provides a novel way to combine low resolution input images with high-resolution ground-truths efficiently. 
Our architecture adds multiple up-scaling layers with skip connections at the end of a U-Net-like architecture to leverage the power of high-resolution ground truths with minimal compromises for the lightweight nature of low-resolution inputs. 
Our architectures up-scaling layers can be added to any conventional U-Net-like network without making any changes to the base network.  
We showed that the predictions generated by our architecture significantly improve the performance on the Decathlon and BraTs datasets.
This shows that our architecture is a valuable alternative to maintain prediction quality if resolution reduction is necessary because of hardware limitations. 
%Our architecture is open-source to ensure reproducible research and accessibility. 
%The source code is online at \url{https://BlindedLinkForReview}. 
%\section*{Acknowledgment}

%\section*{Acknowledgments}
%This work is part of the Moore4Medical project funded by the ECSEL Joint Undertaking under grant number H2020-ECSEL-2019-IA-876190.
%This work was also supported in parts by the NYUAD’s Research Enhancement Fund (REF) Award on “eDLAuto: An Automated Framework for Energy-Efficient Embedded Deep Learning in Autonomous Systems”, and by the NYUAD Center for Artificial Intelligence and Robotics (CAIR), funded by Tamkeen under the NYUAD Research Institute Award CG010.

\bibliography{our_framework}

\begin{thebibliography}{10}

\bibitem{EEG2}
S.~L. Oh, Y.~Hagiwara, U.~Raghavendra, R.~Yuvaraj, N.~Arunkumar, M.~Murugappan, and U.~R. Acharya, ``A deep learning approach for parkinson’s disease diagnosis from eeg signals,'' {\em Neural Comput. and App.}, vol.~32, pp.~10927--10933, 2020.

\bibitem{EEG3}
L.~A. Gemein, R.~T. Schirrmeister, P.~Chrabaszcz, D.~Wilson, J.~Boedecker, A.~Schulze-Bonhage, F.~Hutter, and T.~Ball, ``Machine-learning-based diagnostics of eeg pathology,'' {\em NeuroImage}, vol.~220, p.~117021, 2020.

\bibitem{SR2}
E.~M. Masutani, N.~Bahrami, and A.~Hsiao, ``Deep learning single-frame and multiframe super-resolution for cardiac mri,'' {\em Radiology}, vol.~295, no.~3, pp.~552--561, 2020.

\bibitem{SR3}
S.~Zhang, G.~Liang, S.~Pan, and L.~Zheng, ``A fast medical image super resolution method based on deep learning network,'' {\em IEEE Access}, vol.~7, pp.~12319--12327, 2018.

\bibitem{CADC1}
K.~Suzuki, ``Pixel-based machine learning in medical imaging,'' {\em Journal of Biomed. Imaging}, vol.~2012, pp.~1--1, 2012.

\bibitem{CADC2}
J.~Antony, K.~McGuinness, N.~E. O'Connor, and K.~Moran, ``Quantifying radiographic knee osteoarthritis severity using deep convolutional neural networks,'' in {\em Inter. conf. on pattern recog.}, pp.~1195--1200, IEEE, 2016.

\bibitem{CADC3}
W.~Shen, M.~Zhou, F.~Yang, C.~Yang, and J.~Tian, ``Multi-scale convolutional neural networks for lung nodule classification,'' in {\em Inter. Conf. for Inf. Proc. in Med. Imaging}, pp.~588--599, Springer, 2015.

\bibitem{CADOD1}
C.-W. Wang, C.-T. Huang, M.-C. Hsieh, C.-H. Li, S.-W. Chang, W.-C. Li, R.~Vandaele, R.~Mar{\'e}e, S.~Jodogne, P.~Geurts, {\em et~al.}, ``Evaluation and comparison of anatomical landmark detection methods for cephalometric x-ray images: a grand challenge,'' {\em IEEE trans. on medical imaging}, vol.~34, no.~9, pp.~1890--1900, 2015.

\bibitem{CADOD2}
B.~D. De~Vos, J.~M. Wolterink, P.~A. De~Jong, M.~A. Viergever, and I.~I{\v{s}}gum, ``2d image classification for 3d anatomy localization: employing deep convolutional neural networks,'' in {\em Medical imaging 2016: Image proc.}, vol.~9784, pp.~517--523, SPIE, 2016.

\bibitem{CADOD3}
Z.~Li, M.~Dong, S.~Wen, X.~Hu, P.~Zhou, and Z.~Zeng, ``Clu-cnns: Object detection for medical images,'' {\em Neurocomput.}, vol.~350, pp.~53--59, 2019.

\bibitem{CADSS1}
M.~Marsousi, K.~N. Plataniotis, and S.~Stergiopoulos, ``Shape-based kidney detection and segmentation in three-dimensional abdominal ultrasound images,'' in {\em Int. Conf. of the IEEE Engineering in Med. and Bio. Society}, pp.~2890--2894, 2014.

\bibitem{CADSS2}
{\"O}.~{\c{C}}i{\c{c}}ek, A.~Abdulkadir, S.~S. Lienkamp, T.~Brox, and O.~Ronneberger, ``3d u-net: learning dense volumetric segmentation from sparse annotation,'' in {\em Med. Image Comput. and Comput.-Ass. Interv. (MICCAI)}, pp.~424--432, Springer, 2016.

\bibitem{CADSS3}
M.~Rezaei, H.~Yang, and C.~Meinel, ``Recurrent generative adversarial network for learning imbalanced medical image semantic segmentation,'' {\em Multimed. Tools and Appl.}, vol.~79, no.~21-22, pp.~15329--15348, 2020.

\bibitem{UNET}
O.~Ronneberger, P.~Fischer, and T.~Brox, ``U-net: Convolutional networks for biomedical image segmentation,'' in {\em Med. Image Comput. and Comput.-Ass. Interv.}, pp.~234--241, Springer, 2015.

\bibitem{ELUNET}
Y.~Deng, Y.~Hou, J.~Yan, and D.~Zeng, ``Elu-net: an efficient and lightweight u-net for medical image segmentation,'' {\em IEEE Access}, vol.~10, pp.~35932--35941, 2022.

\bibitem{UNET3+}
H.~Huang, L.~Lin, R.~Tong, H.~Hu, Q.~Zhang, Y.~Iwamoto, X.~Han, Y.-W. Chen, and J.~Wu, ``Unet 3+: A full-scale connected unet for medical image segmentation,'' in {\em IEEE inter. conf. on acoustics, speech and signal processing (ICASSP)}, pp.~1055--1059, 2020.

\bibitem{DECA}
M.~Antonelli, A.~Reinke, S.~Bakas, K.~Farahani, A.~Kopp-Schneider, B.~A. Landman, G.~Litjens, B.~Menze, O.~Ronneberger, R.~M. Summers, {\em et~al.}, ``The medical segmentation decathlon,'' {\em Nature comm.}, vol.~13, no.~1, p.~4128, 2022.

\bibitem{BRATS1}
B.~H. Menze, A.~Jakab, S.~Bauer, J.~Kalpathy-Cramer, K.~Farahani, J.~Kirby, Y.~Burren, N.~Porz, J.~Slotboom, R.~Wiest, {\em et~al.}, ``The multimodal brain tumor image segmentation benchmark (brats),'' {\em IEEE trans. on med. imaging}, vol.~34, no.~10, pp.~1993--2024, 2014.

\bibitem{BRATS2}
S.~Bakas, H.~Akbari, A.~Sotiras, M.~Bilello, M.~Rozycki, J.~S. Kirby, J.~B. Freymann, K.~Farahani, and C.~Davatzikos, ``Advancing the cancer genome atlas glioma mri collections with expert segmentation labels and radiomic features,'' {\em Scien. data}, vol.~4, no.~1, pp.~1--13, 2017.

\bibitem{BRATS3}
S.~Bakas, M.~Reyes, A.~Jakab, S.~Bauer, M.~Rempfler, A.~Crimi, R.~T. Shinohara, C.~Berger, S.~M. Ha, M.~Rozycki, {\em et~al.}, ``Identifying the best machine learning algorithms for brain tumor segmentation, progression assessment, and overall survival prediction in the brats challenge,'' {\em arXiv preprint:1811.02629}, 2018.

\bibitem{RESNET}
C.~Szegedy, S.~Ioffe, V.~Vanhoucke, and A.~Alemi, ``Inception-v4, inception-resnet and the impact of residual connections on learning,'' in {\em AAAI conf. on arti. Intell.}, vol.~31, 2017.

\bibitem{tan2021multimodal}
L.~Tan, W.~Ma, J.~Xia, and S.~Sarker, ``Multimodal magnetic resonance image brain tumor segmentation based on acu-net network,'' {\em IEEE Access}, vol.~9, pp.~14608--14618, 2021.

\bibitem{myronenko20193d}
A.~Myronenko, ``3d mri brain tumor segmentation using autoencoder regularization,'' in {\em Inter. MICCAI Brainlesion Works.}, pp.~311--320, Springer, 2019.

\bibitem{ahmad2021mh}
P.~Ahmad, H.~Jin, R.~Alroobaea, S.~Qamar, R.~Zheng, F.~Alnajjar, and F.~Aboudi, ``Mh unet: A multi-scale hierarchical based architecture for medical image segmentation,'' {\em IEEE Access}, vol.~9, pp.~148384--148408, 2021.

\bibitem{jiang2020two}
Z.~Jiang, C.~Ding, M.~Liu, and D.~Tao, ``Two-stage cascaded u-net: 1st place solution to brats challenge 2019 segmentation task,'' in {\em Inter. MICCAI Brainlesion Works.}, pp.~231--241, Springer, 2020.

\bibitem{bukhari2021e1d3}
S.~T. Bukhari and H.~Mohy-ud Din, ``E1d3 u-net for brain tumor segmentation: Submission to the rsna-asnr-miccai brats 2021 challenge,'' in {\em Inter. MICCAI Brainlesion Works.}, pp.~276--288, Springer, 2021.

\bibitem{peiris2021reciprocal}
H.~Peiris, Z.~Chen, G.~Egan, and M.~Harandi, ``Reciprocal adversarial learning for brain tumor segmentation: a solution to brats challenge 2021 segmentation task,'' in {\em Inter. MICCAI Brainlesion Works.}, pp.~171--181, Springer, 2021.

\bibitem{zabihollahy2019automated}
F.~Zabihollahy, N.~Schieda, S.~Krishna~Jeyaraj, and E.~Ukwatta, ``Automated segmentation of prostate zonal anatomy on t2-weighted (t2w) and apparent diffusion coefficient (adc) map mr images using u-nets,'' {\em Medical physics}, vol.~46, no.~7, pp.~3078--3090, 2019.

\bibitem{nie2019semantic}
D.~Nie and D.~Shen, ``Semantic-guided encoder feature learning for blurry boundary delineation,'' {\em arXiv preprint:1906.04306}, 2019.

\bibitem{zhu2019boundary}
Q.~Zhu, B.~Du, and P.~Yan, ``Boundary-weighted domain adaptive neural network for prostate mr image segmentation,'' {\em IEEE trans. on medical imaging}, vol.~39, no.~3, pp.~753--763, 2019.

\bibitem{wang2019deeply}
B.~Wang, Y.~Lei, S.~Tian, T.~Wang, Y.~Liu, P.~Patel, A.~B. Jani, H.~Mao, W.~J. Curran, T.~Liu, {\em et~al.}, ``Deeply supervised 3d fully convolutional networks with group dilated convolution for automatic mri prostate segmentation,'' {\em Medical physics}, vol.~46, no.~4, pp.~1707--1718, 2019.

\bibitem{UNET++}
Z.~Zhou, M.~M. Rahman~Siddiquee, N.~Tajbakhsh, and J.~Liang, ``Unet++: A nested u-net architecture for medical image segmentation,'' in {\em Inter. Works., DLMIA 2018, and ML-CDS 2018, with MICCAI}, pp.~3--11, Springer, 2018.

\end{thebibliography}
\bibliographystyle{ieeetr}

%\end{thebibliography}
\vspace{12pt}
\color{red}

\end{document}